\documentclass[a4paper,11pt]{article}
\pdfoutput=1 

\usepackage{jcappub} 
\usepackage{comment}
\usepackage{verbatim}
\usepackage{url}

\usepackage[T1]{fontenc} 

\title{Neutrino pair annihilation ($\nu{\bar \nu}\to e^-e^+$) in the presence of quintessence  surrounding a black hole}

\author[a,b]{G. Lambiase,}
\author[a,b,1]{L. Mastrototaro.\note{Corresponding author.}}

\affiliation[a]{Dipartimento di Fisica ``E.R Caianiello'', Università degli Studi di Salerno,\\ Via Giovanni Paolo II, 132 - 84084 Fisciano (SA), Italy}
\affiliation[b]{ Istituto Nazionale di Fisica Nucleare - Gruppo Collegato di Salerno - Sezione di Napoli,\\ Via Giovanni Paolo II, 132 - 84084 Fisciano (SA), Italy}

\emailAdd{lambiase@sa.infn.it}
\emailAdd{lmastrototaro@unisa.it}

\abstract{Quintessence fields, introduced to explain the speed-up of the Universe, might affect the geometry of spacetime surrounding black holes, as compared to the standard Schwarzschild and Kerr geometries. In this framework, we study the neutrino pairs annihilation into electron-positron pairs ($\nu{\bar \nu}\to e^-e^+$) near the surface of a neutron star, focusing, in particular, on the Schwarzschild-like geometry in presence of quintessence fields. The effect of the latter is to increase the photon-sphere radius ($R_{ph}$), increasing in such a way the maximum energy deposition rate near to $R_{ph}$. The rate turns out to be several orders of magnitude greater than the rate computed in the framework of General Relativity. These results might provide a rising in the GRBs energy emitted from a close binary neutron star system and might be used to constraints the parameters of the quintessence model. Finally we theoretically study the effects of rotation on the neutrino energy deposition}
\begin{document}
\maketitle
\flushbottom

\section{Introduction}
\label{intro}

Recent cosmological observations suggest that our Universe is currently undergoing to an accelerated expansion \cite{riess,riess1,riess2,riess3,riess4,riess5}. To explain such a phase of the Universe evolution, 
several {\it alternative} or {\it modified} theories of gravity have been proposed, which allow, among the other things, to encompass several shortcomings of the cosmology based on General Relativity, that is, the standard cosmological model.  For example, higher-order curvature invariants allow to get inflationary behaviour, as well as to explain the flatness and horizon problems \cite{starobinski,starobinski1} (for further applications and different models, see Refs. \cite{Capozziello:2011et,Tino:2020nla,cosmo1,cosmo2,cosmo3,cosmo4,cosmo5,cosmo6,cosmo7,cosmo8,cosmo9,cosmo10,cosmo11,cosmo12,cosmo13,cosmo14,cosmo15,odi,cosmo17,cosmo18,cosmo19,cosmo20,cosmo21,cosmo22,cosmo23,cosmo24,cosmo25,Nojiri:2017ncd,Benetti:2020hxp,Bernal:2020ywq}). This approach, as well as ones related to it, follows from the fact that the high curvature regime requires that curvature invariants are necessary for building up self-consistent effective actions in curved spacetime \cite{birrell,shapiro,barth}.

Another possibility to explain the speed-up of the present Universe is to invoke some exotic component, called dark energy. Despite the observational evidence, the nature and origin of dark energy is till now a source of vivid debate. A candidate of Dark Energy could be the Quintessence \cite{Jamil:2014rsa}, which is characterized by the fact that it may generate a negative pressure, and, since it exists everywhere in the Universe, it can cause the observed accelerated phase.

The existence of a quintessence field diffuse in the Universe has suggested the possibility that it could be present around a huge massive gravitational object, with the consequence that the spacetime around it gets deformed. Such a possibility has been studied by Kiselev in Ref. \cite{Kiselev:2002dx}. 
In this work, the author solved the Einstein field equations for static spherically symmetric quintessence surrounding a black hole in $d=4$ dimensions. Here it is also shown that
appropriate conditions 
of the energy-momentum tensor allow inferring the correct limit to the known solutions corresponding to the electromagnetic static field and the cosmological constant.  
The Kiselev solution is characterized by two parameters: the parameter of state $ w_q$, that is bounded as $-1< w_q<-\frac{1}{3}$, and the quintessence parameter $c$. The generalization of these results to $d$ dimensions has been studied in \cite{Chen:2008ra}. The studies of quintessential black holes are also motivated from M-theory/superstring inspired models \cite{Belhaj:2020oun,HeydarFard:2007bb,HeydariFard:2007qs}. For applications see Refs. \cite{Toshmatov:2015npp,Abdujabbarov:2015pqp,Ghosh:2015ovj,Jamil:2014rsa,Belhaj:2020rdb,Khan5essence,Abbas:2019olp,Khan:2020ngg,Javed:2019jag,Khan:2020pgl,Uniyal:2014paa}.

The aim of this paper is to investigate the effects of the presence of the quintessence field around a gravitational object described by the black hole metric on the neutrino pair annihilation efficiency\footnote{The  propagation of neutrino in gravitational fields has been studied both in GR~ \cite{nuGR,nugr1,nugr2,nugr3,nugr4,nugr5,nugr6,Dvornikov:2021hps,nugr7,nugr8,nugr9,MosqueraCuesta:2017iln,Cuesta:2008te,Lambiase:2004qk} and in modified gravity \cite{nuETG,nuetg1,nuetg2}.} ($\nu\bar{\nu}\rightarrow e^+e^-$). Moreover, considering the merging of a binary system of neutron star, the $e^+ e^-$ in turn generate gamma ray that may provide a possible explanation of the observed GRBs~\cite{Salmonson:1999es}. 
The studies of the energy source of GRBs induced by the neutrino-antineutrino annihilation into electrons and positrons, and their relevance in physics and astrophysics, have a quite long history \cite{Co86,Co87,Goodman:1986we,Salmonson:1999es,Salmonson:2001tz,Asano:2000ib,Asano:2000dq,Prasanna:2001ie,Birkl:2006mu,George:2002xs,Pa90,Meszaros:1992gc,Ruffert:1998qg,Asano:2002qz}. 
It is worth to mention that the neutrino pair annihilation rate into electron pairs between two neutron stars in a binary system has been investigated in \cite{Salmonson:2001tz}, while the study of the structure of neutron star disks based on the two-region (inner and outer) disk scenario has been performed in \cite{Zhang:2009ew}, where the neutrino annihilation luminosity from the disk has been calculated in various cases (see also \cite{Mallick:2008iv,Bhattacharyya:2009nm,Chan:2009mw,Kovacs:2009dv,Kovacs:2010zp}). 
The neutrino-antineutrino annihilation into electron-positron pairs near the surface of a neutron star assuming that the gravitational background is described by modified theories of gravity has been investigated in \cite{Lambiase:2020iul}.

Motivated by the fact that gravity may influence energy deposition rate near to the photon-sphere, $R_{ph}$, as compared to one derived in the framework of Newtonian gravity \cite{Salmonson:1999es}, we extend this analysis to the case of black hole spacetimes modified by quintessence fields. In particular, we focus on the process $\nu\bar{\nu}\rightarrow e^+e^-$, which is important also for the delay shock mechanism into the Type II Supernova:  indeed, at late time, from the hot proto-neutron star, the energy is deposited into the supernova envelope via neutrino pair annihilation and neutrino-lepton scattering \cite{Lambiase:2020iul}. As we shall see, the computation of the efficiency of the process $\nu\bar\nu\rightarrow e^+e^-$  may increase by more than one order of magnitude respect to GR, making possible to explain the huge energy of the GRB energy emitted in the merging of a binary neutron star system, considering the neutrino annihilation as the unique source. Indeed electron pairs subsequently annihilate into photons (fireball model \cite{Leng:2014dfa}) that ultimately powers the GRB. We have approximate the effect considering that all the energy released into electrons is converted into photons~\cite{Salmonson:1999es} and considering the escape probability for deposited energy equal to its minimum value of $1/2$ at the neutrinosphere (see App.~\ref{appendice}).

\vspace{0.1in}

The paper is organized as follows. In the next section, we recall the main features related to neutrino energy deposition. In particular, we write the total energy deposition and luminosity induced by neutrino annihilation processes. In Section 3 we discuss the effects of quintessential matter surrounding gravitational sources on the neutrino energy deposition. Such theoretical analysis could provides also an enhancement of the GRB energy emitted from the merging of a binary neutron system and lead to constraints on the quintessence model. We will derive general results about the energy deposition rate for the model of neutrinos spherically emitted from the neutrinosphere as done in Ref.~\cite{Salmonson:1999es}.
\footnote{The relativistic effects on the energy deposition rate generated by neutrinos emitted from the accretion disk around a rotating black hole described by modified theories of gravity will be faced elsewhere.}.
In the last section, we discuss our conclusions.   

\section{Energy deposition rate  by the neutrino annihilation process}
\label{Geodetic}

In this section we recall the main features to treat the energy deposition in curved spacetimes  \cite{Prasanna:2001ie,Lambiase:2020iul}.  As mentioned in the Introduction, previous calculations of the $\nu {\bar \nu}\to e^- e^+$ reaction in the vicinity of a neutron star have been first based on Newtonian gravity \cite{Co86,Co87}, then the effect of gravity has been incorporated for static stars \cite{Salmonson:1999es,Salmonson:2001tz}, and then extended to
rotating stars \cite{Prasanna:2001ie,Bhattacharyya:2009nm}. 

We consider the spacetime around a black hole in presence of the quintessence field. The solution of Einstein's field equations for a static spherically symmetric quintessence surrounding a black hole in 4 dimension is described by the diagonal metric \cite{Kiselev:2002dx,Chen:2008ra} (see also  \cite{Belhaj:2020oun,HeydarFard:2007bb,HeydariFard:2007qs} for  quintessential black holes  motivated from M-theory/superstring inspired models)
  \begin{equation}
g_{\mu\nu}=\text{diag}\left(-f(r), f^{-1}(r), r^2, r^2 \sin^2 \theta\right)\,,
\label{metric}
\end{equation}
with
\begin{equation}\label{frmetric}
f(r)=1-\frac{2M}{r}-\frac{c}{r^{3\omega_q+1}}\,.
\end{equation}
See Appendix B for some details.


The energy deposition per unit time and per volume is given by (we shall use natural units, $c=\hbar =1$)  \cite{Goodman:1986we} 
\begin{equation}\label{qdotgeneral}
\dot{q}(r) = \iint f_\nu({\bf p}_\nu,r)
f_{\overline{\nu}}({\bf p}_{\overline{\nu}},r)
 \left[\sigma |{\bf v}_\nu - {\bf v}_{\overline{\nu}} | \varepsilon_\nu
\varepsilon_{\overline{\nu}} \right] 
\frac{ \varepsilon_\nu + \varepsilon_{\overline{\nu}}}
{
 \varepsilon_\nu \varepsilon_{\overline{\nu}} } d^3{\bf p}_\nu d^3{\bf p}_{\overline{\nu}}\,,
\end{equation}
where $f_{\nu,{\overline{\nu}}}$ are the neutrino number densities in phase space, ${\bf v}_\nu$ the neutrino velocity, and $\sigma$ is the rest frame cross section.  
Since the term $\sigma |{\bf v}_\nu - {\bf v}_{\overline{\nu}} | \varepsilon_\nu
\varepsilon_{\overline{\nu}}$ is Lorentz invariant, it can be calculated in the center-of-mass frame, and turns out to be
\begin{equation}
\sigma |{\bf v}_\nu - {\bf v}_{\overline{\nu}} | \varepsilon_\nu
\varepsilon_{\overline{\nu}}  = \frac{D G^2_F}{3\pi} (\varepsilon_\nu
\varepsilon_{\overline{\nu}} - {\bf p}_\nu \cdot
{\bf p}_{\overline{\nu}} c^2 )^2\,,
\end{equation}
where $G_F=5.29 \times 10^{-44}$ cm$^2$ MeV$^{-2}$ is the Fermi constant,
\begin{equation}
D=1\pm4\sin^2\theta_W+8\sin^4\theta_W, 
\end{equation}
$\sin^2\theta_W=0.23$ is the Weinberg angle, and the plus sign is for electron neutrinos and antineutrinos while the minus sign is for muon and tau type. $T(r)$ is the temperature measured by the local observer and $\Theta(r)$ is the angular integration factor. In these frameworks the energy of the neutrinos
is $\gtrsim 10$ MeV, so that the mass of the electrons can be neglected. It is possible to obtain that the general expression of the rate per unit time and unit volume of the $\nu\bar\nu\rightarrow e^+e^-$ process is \cite{Salmonson:1999es} 
\begin{equation}\label{qpunto}
\dot{q}=\frac{7DG_F^2\pi^3\xi(5)}{2}[k T(r)]^9\Theta(r) \,.
\end{equation}
 The evaluation of $T(r)$ and $\Theta(r)$ account for the gravitational redshift and path bending.

 To write the latter in terms of observed luminosity $L_\infty$, which is an observable quantity, one has to has to express the temperature of the free streaming neutrinos at radius $r$ in terms of their temperature at the neutrinosphere radius
$R$ using the appropriate gravitational red-shift.  Temperature, like energy, varies linearly with red-shift. Following the procedure of Ref. \cite{Salmonson:1999es}, one finds that
the above expressions, written in terms of the metric component $g_{00}=g_{rr}^{-1}=f(r)$, are given by (see for example \cite{Lambiase:2020iul})
\begin{align}
    \Theta(r)&=\frac{2\pi^3}{3}(1-x)^4(x^2+4x+5) \,, \\
    T(r)&=\frac{\sqrt{f(R)}}{\sqrt{f(r)}}T(R) \,, \\
    L_{\infty}&=f(R)L(R) \,, \\
    L(R) & = L_\nu + L_{\overline{\nu}} = \frac{7}{4} \, a \pi R^2 T^4(R) \,.
\end{align}
Here $x=\sin^2\theta_r$, where $\theta_r$ is the angle between the trajectory and the tangent velocity in terms of local radial and longitudinal velocities~\cite{Prasanna:2001ie}, and is defined as \cite{Lambiase:2020iul}
\begin{equation}\label{cosr}
\cos\theta_r=\frac{R}{r}\sqrt{\frac{f(r)}{f(R)}},
\end{equation}
$R$ the neutrinosphere radius (the spherical surface where the stellar material is transparent to neutrinos and from which neutrinos are emitted freely), $L(R)$ is the neutrino luminosity at distance $R$, and $a$ the radiation constant. This relation comes from the fact that the impact parameter $b$ is constant on all the trajectory and is related to $\cos\theta_r$ by the relation
\begin{equation}\label{bSchw}
    b=\left(\frac{f(r)}{r\cos\theta_r}\right)^{-1} \,\ .
\end{equation}
The fact that $b$ is constant along the trajectory, implies the existence of a photosphere radius $R_{ph}$ below which a massless particle can not be emitted tangent to the stellar surface. The present discussion is therefore restricted to $R>R_{ph}$. The neutrino emission properties hence mainly depend on the geometry of spacetime. From the equation of the velocities 
\[
\dot{r}^2 = \left(E\dot{t}-L\dot{\phi}\right)f(r)\,, \quad
\dot{\phi}=\frac{L}{r^2}, \quad
\dot{t} =-\frac{E}{f(r)},
 \]
where $E$ and $L$ are the energy and angular momentum at the infinity, one gets the effective potential $V_{eff}$, such that the photonsphere radius follows from the condition $\frac{\partial V_{\mathrm{eff}}}{\partial r}=0$.
The circular orbit is derived imposing $\dot{r}^2=0$.
We note en passant that these results reduce to ones derived in the case of the Schwarzschild geometry, $R_{ph}=3M$, as calculated in \cite{Salmonson:1999es}.

The integration of $\dot{q}$ from $R$ to infinity gives the total amount of local energy deposited by the neutrino annihilation process (for a single neutrino flavour) for time units
\begin{equation}\label{Qdot}
    \dot{Q}= 4\pi \int_R^{\infty}  dr \frac{r^2}{\sqrt{f(r)}}\, \dot{q} \,,
\end{equation}
which can be cast cast in the form
\begin{equation}
\dot{Q}_{51}=1.09\times 10^{-5}\mathcal{F}\left(\frac{M}{R}\right)DL_{51}^{9/4}R_6^{-3/2} \,,
\label{contoQ}
\end{equation}
where $\dot{Q}_{51}$ and $L_{51}$ are the total energy deposition and luminosity, respectively, in units of $10^{51}~\mathrm{erg/sec}$, 
\begin{equation}\label{QL51}
\dot{Q}_{51}\equiv \frac{{\dot Q}}{10^{51}\text{erg/sec}}\,, \quad L_{51}\equiv \frac{L_{\infty}}{10^{51}\text{erg/sec}}\,,
\end{equation}
$R_6$ is the neutrinosphere radius in units of $10~\mathrm{km}$, $R_6\equiv R/10\text{km}$, and
\begin{equation}\label{fMRSchw}
\mathcal{F}\left(\frac{M}{R}\right) = 3 f^{9/4}(R)
\int_1^{\mathrm{R_{ch}}}(x-1)^4(x^2+4x+5)\frac{y^2dy}{f^5(yR)}\,,
\end{equation}
with $y=r/R$.  In the Newtonian limit ($M\rightarrow 0$) one gets $\dot{Q}\to \dot{Q}_{\mathrm{Newt}}$, where $\dot{Q}_{\mathrm{Newt}}$ is obtained from (\ref{contoQ}) (or (\ref{QL51})) with  
$\mathcal{F}(0)=1$. It turns out convenient to define the ratio $\dot{Q}/\dot{Q}_{\mathrm{Newt}}$, so that one gets
 \begin{equation}\label{Qratio}
\frac{\dot{Q}}{\dot{Q}_{\mathrm{Newt}}}=
\mathcal{F}\left(\frac{M}{R}\right)\,.
\end{equation}
In the next Section we will study the ratio (\ref{Qratio}) for astrophysical objects surrounding by Dark Energy quintessence.

\section{Neutrino deposition in BH spacetimes with quintessence}

In this section, we analyze the neutrino pair annihilation into electron-positron pairs near the surface of a gravitational source taking into account the deviation from Schwarzschild solution induced by the presence of the quintessence field. The effect is studied in the regime of strong gravitational field.
In the Schwarzschild geometry with quintessence field, the parameter $\omega_q$ assumes, as stated in the Introduction, the values in the range
 \[ 
 \omega_q \in\,\, \Big]-1,-\frac{1}{3}\Big[ 
 \]
%
The quintessence parameter is constrained by the fact that increasing $c$, the model passes from describing a black hole with an event horizon to one representing a naked singularity. This latter case is represented with the blue curve in Fig.~\ref{f(r)}. Moreover, it can be seen that increasing $c$ the curve changes until the formation of a naked singularity occurs. According to these results, we confine ourselves to values of the parameter $c$ for which the event horizon does exist. 
In what follows we analyzed some specific cases for different values of the parameter quintessence $\omega_q$:

\begin{description}
\item[$\bullet$] $\omega_q = -0.9$ - In Fig. \ref{09}, it is possible to notice that we obtain only an increment of a factor $2$ respect to GR in correspondence of $c=1\times 10^{-2}$. We can notice also that the different behaviour respect to GR for large value of $R/M$ is due to the existence of the horizon at $R/M=13.68$. 
\item[$\bullet$] $\omega_q = -2/3$ - In Fig. \ref{2/3}, it is possible to see that the  relevant enhancement corresponds to $c=5\times 10^{-2}$, up to almost $3$ times with respect to the GR behaviour. As before, the different behaviour respect to GR for large value of $R/M$ is due to the existence of the horizon at $R\sim 16M$.
\item[$\bullet$] $\omega_q = -0.4$ - This case is particularly interesting because, as arises from  Fig. \ref{04}, there is a consistent increment of the fraction of the energy emission rates, up to a factor $\sim 23$ for $c=4\times 10^{-1}$.

\begin{figure}[t]
\centering
  \includegraphics[width=0.9\textwidth]{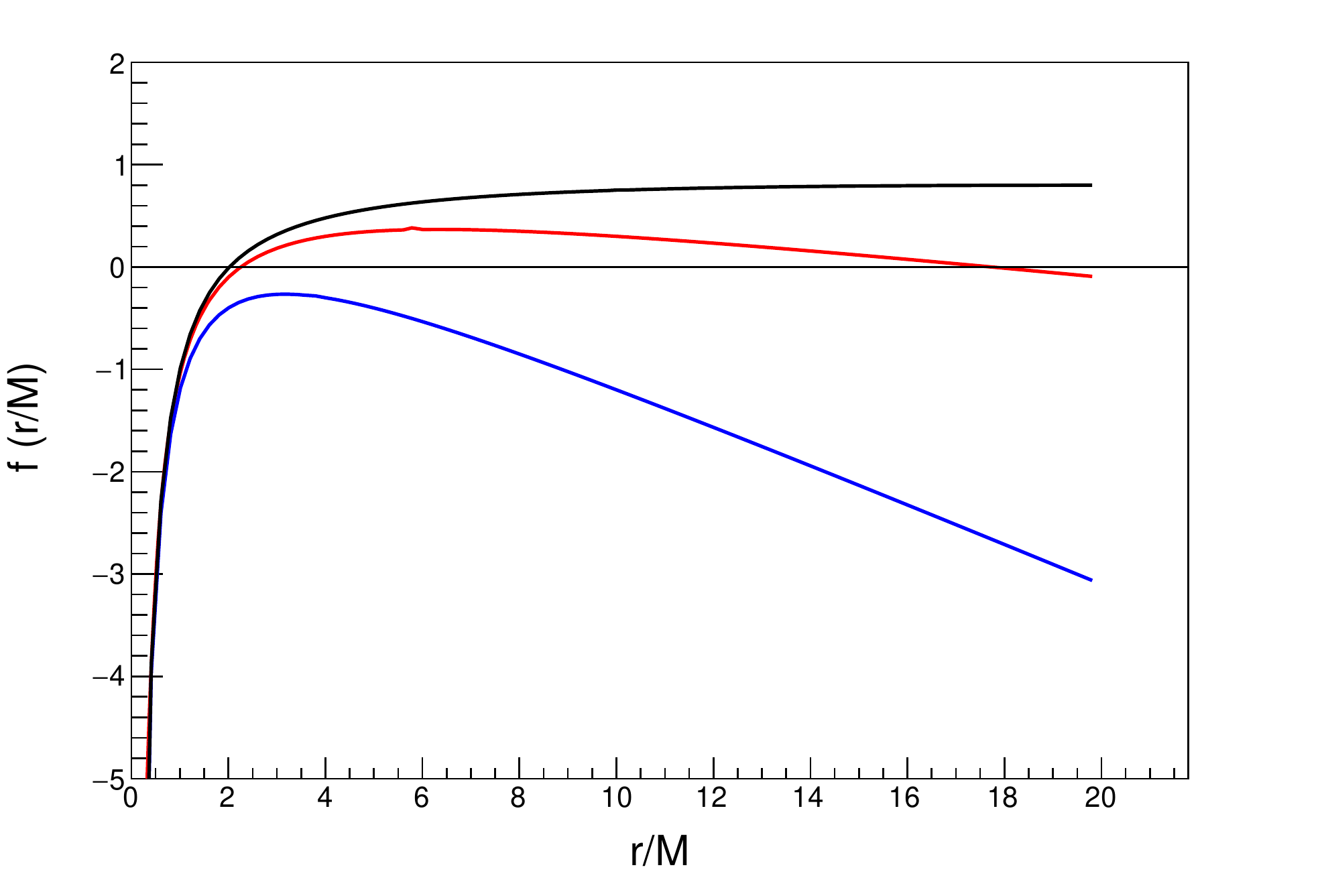}
\caption{The $f(r)$ function defined in (\ref{frmetric}) vs $r/M$, The parameters $\omega_q$ and $c$ are fixed as $\omega_q=-2/3$, $c=0.01$ (black), $c=0.05$ (red) and $c=0.2$ (blue). As arises from the plot, increasing the values of the constant $c$,  the curves approach the event horizon, corresponding to $f(r/M)=0$, until the formation of a naked singularity (blue curve).}
\label{f(r)}       
\end{figure}
\begin{figure}[h!]
\centering
  \includegraphics[width=0.9\textwidth]{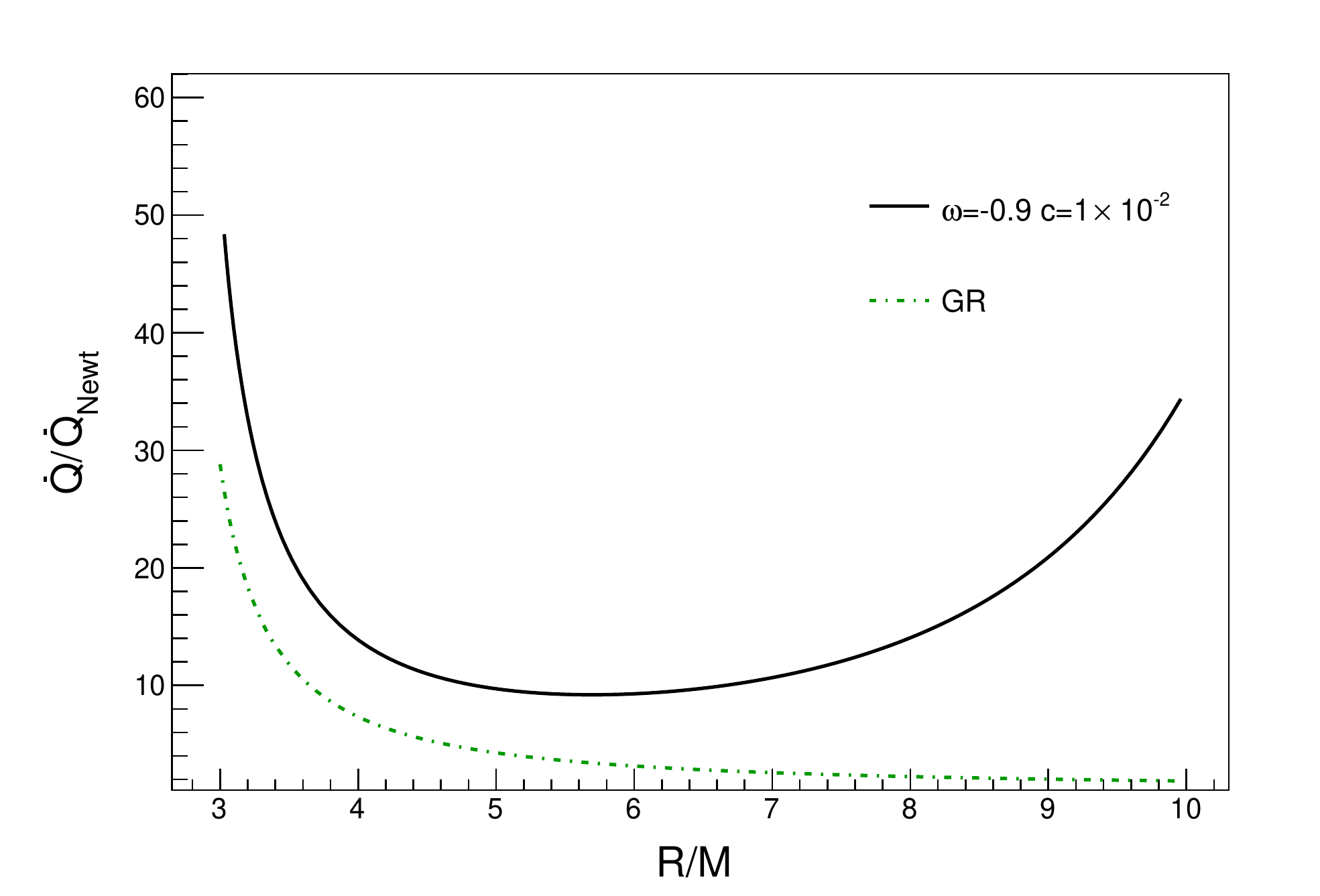}
\caption{Ratio of total energy deposition $\dot{Q}$ for $\omega=-0.9$ to total Newtonian energy deposition $\dot{Q}_{\mathrm{Newt}}$ for two values of the parameter $c$. The green curve shows the GR energy deposition for comparison.}
\label{09}       
\end{figure}
\begin{figure}[h!]
\centering
  \includegraphics[width=0.9\textwidth]{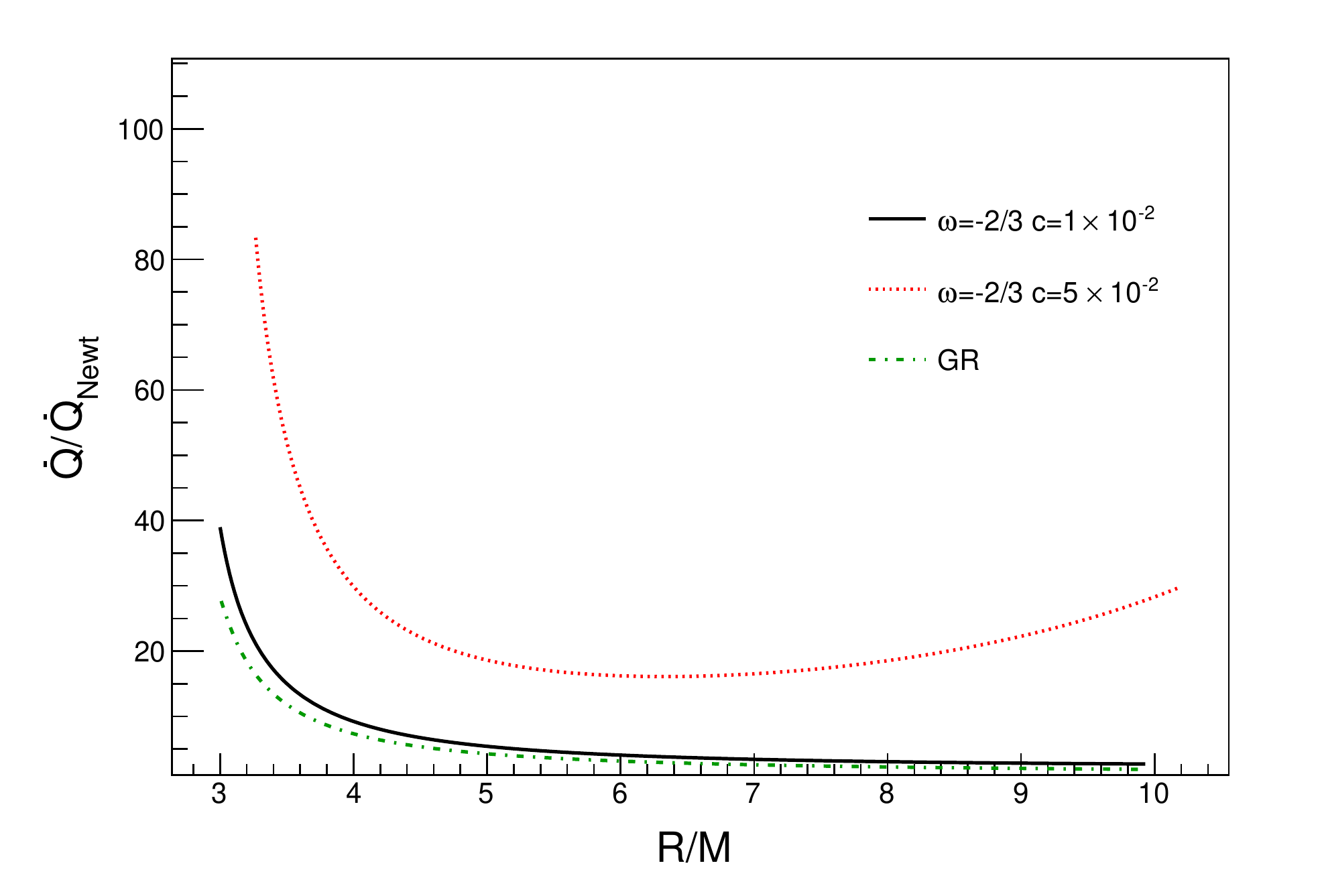}
\caption{Ratio of total energy deposition $\dot{Q}$ for $\omega=-2/3$ to total Newtonian energy deposition $\dot{Q}_{\mathrm{Newt}}$ for three values of the parameter $c$. The green curve shows the GR energy deposition for comparison.}
\label{2/3}       
\end{figure}
\begin{figure}[h!]
\centering
  \includegraphics[width=0.9\textwidth]{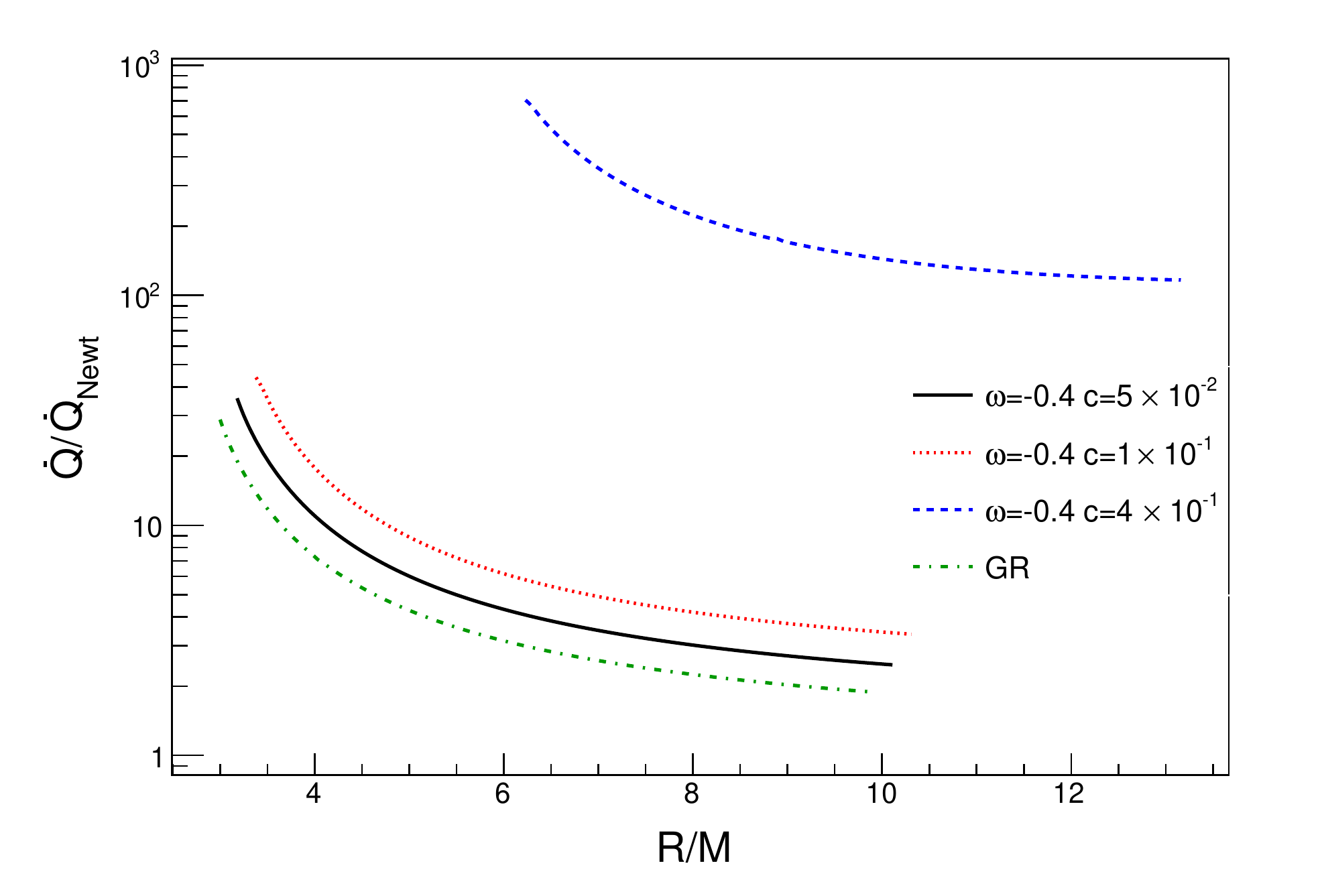}
\caption{Ratio of total energy deposition $\dot{Q}$ for $\omega=-0.4$ to total Newtonian energy deposition $\dot{Q}_{\mathrm{Newt}}$ for three values of the parameter $c$. The green curve shows the GR energy deposition for comparison.}
\label{04}       
\end{figure}

\end{description}

\begin{figure}[h!]
\centering
  \includegraphics[width=0.9\textwidth]{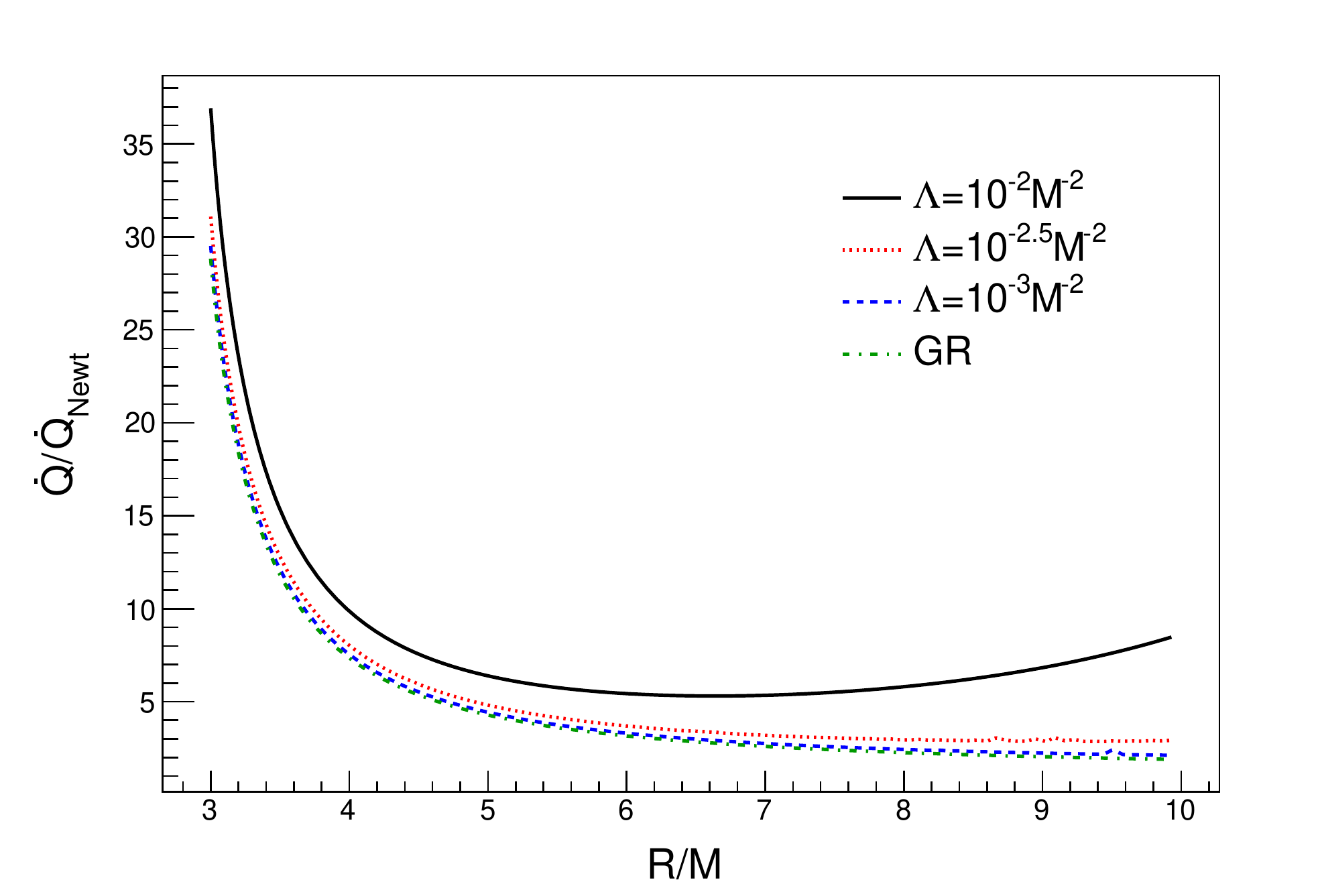}
\caption{Ratio of total energy deposition $\dot{Q}$ for $\Lambda CMD$ model to total Newtonian energy deposition $\dot{Q}_{\mathrm{Newt}}$ for three values of the parameter $\Lambda$. The green curve shows the GR energy deposition for comparison.}
\label{Lambda}       
\end{figure}

For a comparison, we  discuss the effects of a cosmological constant ($\Lambda$CMD model).  The corresponding (gravitational) model is obtained from the metric (\ref{metric}) and (\ref{frmetric}), with $\omega_q=-1$ and $c=\Lambda/3$,
\begin{equation}
    f(r)=1-\frac{2M}{r}-\frac{\Lambda r^2}{3} \,\ .
\end{equation}
Results are shown in Fig.~\ref{Lambda}. As it can be seen, this model does not induce significant differences with respect to GR. 

\subsection{Energy deposition estimations}

The enhancement of the ratio ${\dot Q}/{\dot Q}_{Newt}$ studied in the previous Section, Eq. (\ref{Qratio}), turns out to be relevant for  GRB generation\footnote{Here we just recall the main feature of GRBs. They are classified in two categories, long and short GRBs \cite{Piran:2004ba,Piran:2004qe,Meszaros:2006rc,Piron:2015wtz,Zhang:2011rp}. They are powerful high-energy transient emissions from sources at cosmological distances that are randomly diffuse in the celestial sphere. Moreover, they are characterized by a short phase of intense emission in hard $X$ and $\gamma$ rays, lasting in the range of $\sim {\cal O} (10^{-3})$ sec to $\sim {\cal O} (10^{2})$ sec. 
More specifically, their duration is characterized by the fact that bursts with duration $\lesssim {\cal O}(1)$ sec are referred as short GRBs, bursts with duration $\gtrsim {\cal O}(1)$ sec are referred as long GRBs \cite{Leng:2014dfa}. The firsts are generated by the merging of compact binaries \cite{Eichler:1989ve}, the latter by the collapse of the  Wolf-Rayet star \cite{Woosley:1993wj}. According to the general picture, after following the prompt phase, GRBs exhibit the afterglow phase, a long-lasting activity. During this phase, the observed flux decreases rapidly in time, with an emission peak energy shifting to longer wavelengths ($X$ rays, visible and radio) on time scales that vary from hours to weeks \cite{Piran:2004ba}. The energy rate of short GRBs is
\cite{Piran:2004ba,Piran:2004qe,Meszaros:2006rc,Piron:2015wtz,Zhang:2011rp}
$\dot{E}< 10^{52}\mathrm{erg/s}$. There are several space-mission, which aim is to exploiting GRBs for investigating the early Universe and provide an advancement of multi-messenger and time-domain astrophysics
(see Refs. \cite{Amati:2008hq,Amati:2017npy,Stratta:2017bwq,Gehrels:2004aa,Piron:2011zza} and references therein).}. 
Here we follow the ideas proposed in the papers \cite{Salmonson:2000ke,Salmonson:2002nr,Mathews:1997ts,Salmonson:2001tz,Salmonson:2000bs}. According to them, merging neutron stars in a close binary system undergo to a relativistic
compression and heating over a period of seconds, and may released gravitational binding energy that can be
converted into internal energy. As a consequence,   
it is possible that
neutrinos (thermally produced) can be emitted slightly before the stars collapse to a BH. The total neutrino luminosity  
is huge, up to $10^{53}$ erg/sec \cite{Mathews:1997vw}.  
Neutrinos emerging from the merged star will deposit energy through the annihilation,  $\nu\overline{\nu}$ , leading to the subsequent formation of electron-positron pairs. As shown by Salmonson and Wilson (SW), strong gravitational fields near the stars will bend the neutrino trajectories, enhancing the annihilation and scattering rates \cite{Salmonson:1999es}. In particular, for  $R/M \sim 3 - 4$ (just before stellar collapse) it follows that $\nu\overline{\nu}$ annihilation deposited energy is enhanced by a factor ${\mathcal{F}}(R/M) \sim {\cal O}(8-28)$ (see below). 
The relevant consequence of SW results is that  the hot electron-positron pair plasma, produced by the $\nu {\bar \nu}$ annihilation, generates in turn a relativistic expanding pair-photon plasma
that becomes optically thin, so that photons can be released. SW found that the energy released  in $\gamma$-rays \cite{Salmonson:2000ke}
is such that the  spectral and temporal properties are consistent with the observed short GRBs. As we will show, the enhancement can increase if the spacetime is deformed by the presence of quintessence field, making the mechanism of GRBs generation more efficient, as compared with GR.

\begin{figure}[h!]
\centering
  \includegraphics[width=0.9\textwidth]{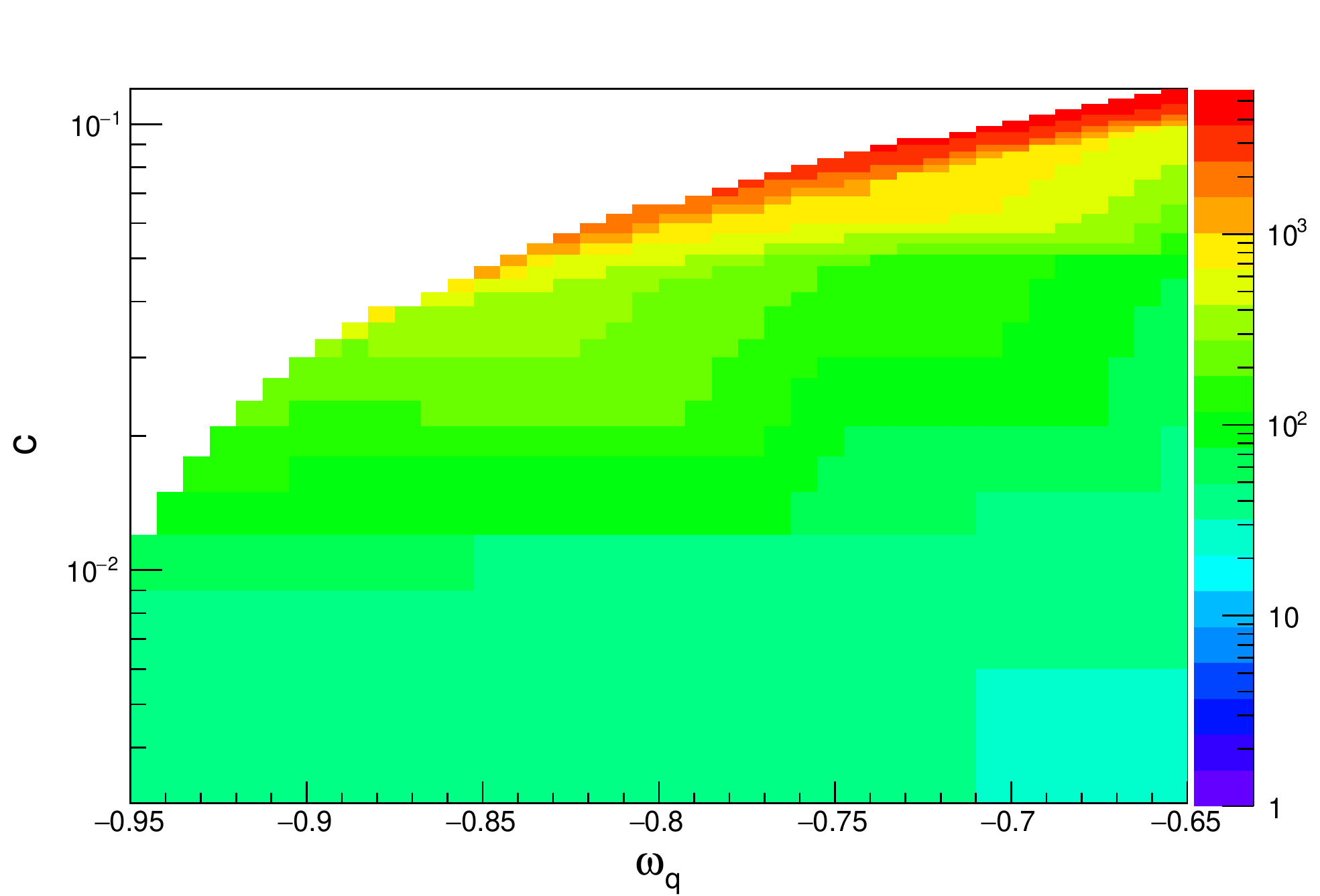}
\caption{Contour plot for $\omega_q\in ]-1,-0.65[$. On the y axis are reported the excluded values of $c$ due to the creation of a naked singularity (white part) and, on the right, the values of $\mathcal{F}(R_{\mathrm{ph}})$. It can be also seen the values of the parameter $c$, for which $\mathcal{F}(R_{\mathrm{ph}}) \sim \mathcal{O}(10^4\textit{-}10^5)$, that are excluded by the energy deposition bounds.}
\label{cont1}       
\end{figure}

\begin{figure}[h!]
\centering
  \includegraphics[width=0.9\textwidth]{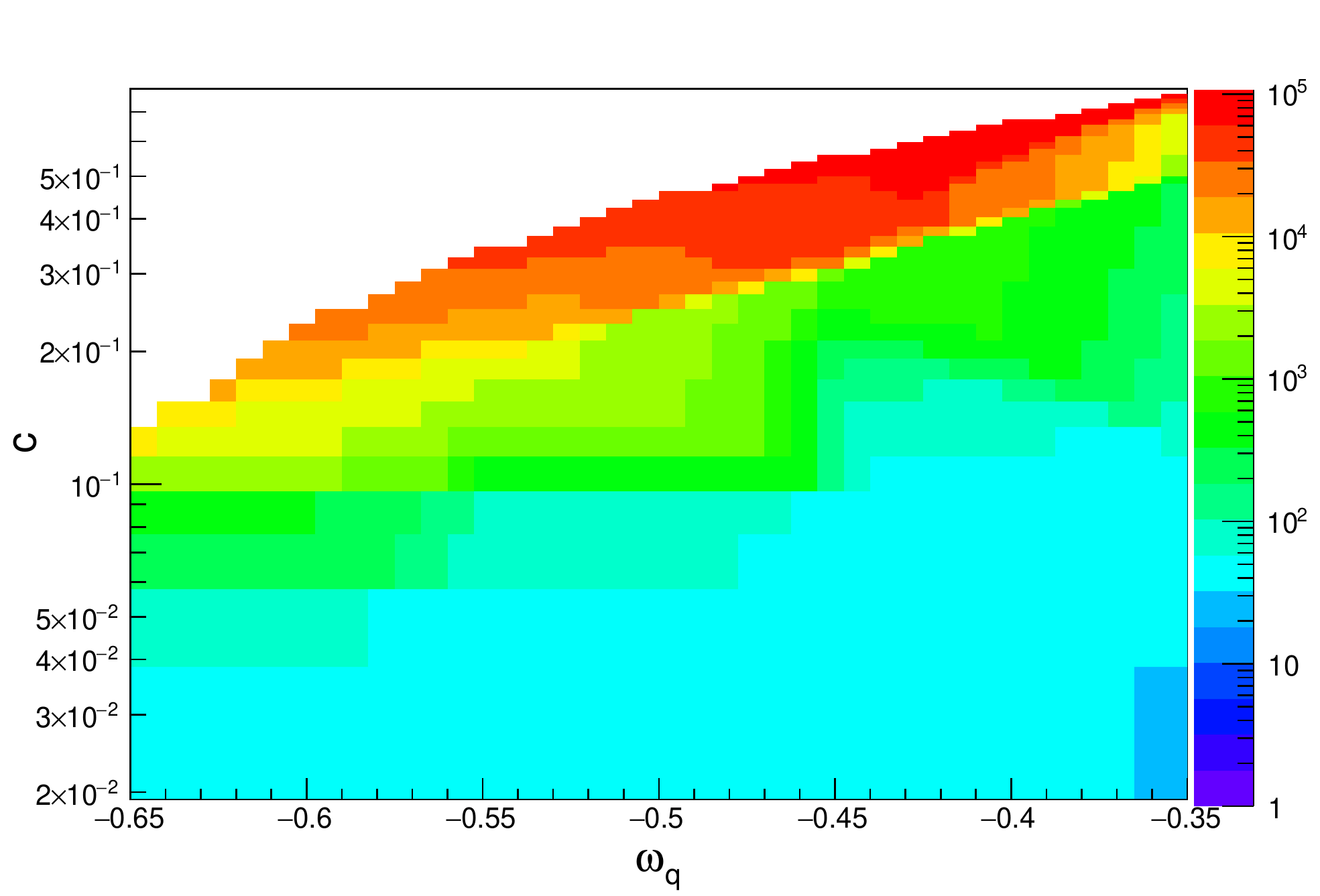}
\caption{Contour plot for $\omega_q\in ]-0.65,-0.35[$. On the y axis are reported the excluded values of $c$ due to the creation of a naked singularity (white part) and, on the right, the values of $\mathcal{F}(R_{\mathrm{ph}})$. It can be also seen the values of the parameter $c$, for which $\mathcal{F}(R_{\mathrm{ph}}) \sim \mathcal{O}(10^4\textit{-}10^5)$, that are excluded by the energy deposition bounds.}
\label{cont2}       
\end{figure}

From Eqs. (\ref{QL51}) and (\ref{contoQ}), integrating in time, it follows that the maximum total energy deposited from the neutrino annihilation process can be explicitly rewritten as  
\begin{equation}
    Q=2.4\times10^{48}~\mathcal{F}(R_{ph})R_6^{-3/2}~ \mathrm{erg}\,\ ,
\end{equation}
where we have considered neutrinos emitted from the photonsphere, a neutrino energy at infinity of $\mathcal{O}(10^{52})\mathrm{erg}$~\cite{Perego:2017fho}\footnote{The final state of a binary neutron star merging is a black hole with a disk. Therefore, due to the used formalism, we have not considered the luminosity that arises from the disk but only that from the merged neutron stars collapsed to a BH.} and $D\sim1.23$.
For a gravitational background described by the Schwarzschild geometry of GR (and $\Lambda$CDM model), considering a radius $R=20~\mathrm{km}$~\cite{Perego:2017fho}, one infers that  the function $\mathcal{F}(R_{ph})$ given by (\ref{fMRSchw}) is of the order $\mathcal{F}(R_{ph})\sim 30$ (we are considering merged neutron stars in a binary system that emits neutrinos slightly before the collapse to a BH), so that the maximum energy reads
\begin{equation}\label{QGR1}
Q_{\mathrm{GR}}\,\,\sim \,\,2.5\times 10^{49}~\mathrm{erg}\,.
\end{equation}
%
%
In the case of Schwarzschild spacetime surrounding by a quintessence field, the function $\mathcal{F}(R_{ph})$  can assume higher values (see Figures \ref{09} - \ref{04}). This implies that the deposited energy is at least two orders of magnitude higher than that in GR.
Considering a maximum true emitted GRB energy of $\mathcal{O}(10^{52}~\mathrm{erg})$ (cfr Tab.2 of Ref.~\cite{Perego:2017fho} \footnote{In Ref~\cite{Perego:2017fho},the authors use the relation $E_{\mathrm{true}}=\left(1-\cos\theta_{\mathrm{jet}}\right)\left(E_{\mathrm{\gamma,iso}}+E_{\mathrm{kin,iso}}\right)$ to determine the true GRB energy emitted related to some GRB jets data of Ref.~\cite{Fong_2015}}), that neutrino pair annihilation is the only source of energy powering short GRBs and that the deposited energy is converted very efficiently to the relativistic jet energy, we infer a constraint on the quintessence model. For the latter case, also considering 
%
the increasing of the photonsphere radius respect to the one derived in  GR, as it can be seen in Fig.~\ref{04}, one obtains that the maximum allowed value of $\mathcal{F}(R_{ph})$ is $\mathcal{O}(10^4)$ .
The contour plots given in Figs.~\ref{cont1} and \ref{cont2} show the value of $\mathcal{F}(R_{\mathrm{ph}})$ for  allowed value of $\omega_q$ and $c$. Therefore, one can infers, for $\mathcal{F}(R_{\mathrm{ph}}) \sim \mathcal{O}(10^4\textit{-}10^5)$, the values of the parameter $c$ that are not allowed in the considered scenario.



\vspace{0.2in}

\subsection{Neutrino deposition in Kerr BH spacetimes with quintessence}

For completeness, we report the results corresponding to the case of a slowly rotating star with quintessence.  To the second order approximation in angular momentum of the source, $J$, the line element  for a particle in motion in the equatorial plane has the form \cite{Toshmatov:2015npp,Ghosh:2015ovj}
\begin{equation}
\begin{split}\label{metricKerr}
ds^2=&-\left(1-\frac{2M}{r}-\frac{2J^2}{r^4}-\frac{c}{r^{3\omega_q+1}}\right)dt^2+\left(1-\frac{2M}{r}+\frac{2J^2}{r^4}-\frac{c}{r^{3\omega_q+1}}\right)^{-1}dr^2\\&+r^2 d\phi^2-\left(\frac{4}{r}+\frac{2c}{r^{3\omega_q+1}}\right)J dt d\phi \,\ .
\end{split}
\end{equation}
Using a similar procedure as in Ref.~\cite{Prasanna:2001ie}, Eqs. (\ref{bSchw}), (\ref{cosr}), and (\ref{fMRSchw}) are generalized as
\begin{align}
    b&=\left[\frac{J}{r}\left(\frac{2}{r}+\frac{c}{r^{3\omega_q+1}}\right)+\frac{(1-2M/r+2J^2/r^4-c/r^{3\omega_q+1})^{1/2}}{r\cos\theta_r}\right] \,\ , \\
    \cos\theta_r&=\frac{R^3r^2(1-2M/r+2J^2/r^4-c/r^{3\omega_q+1})^{1/2}}{J(2+cr^{-3\omega_q})(r^3-R^3)+R^2r^3(1-2M/R+2J^2/R^4-c/R^{3\omega_q+1})^{1/2}} \,\ , 
    \label{cosrJ}\\
    \mathcal{F}_J\left(\frac{M}{R}\right)&= 3\left[g_{00}(R)\right]^{9/4}
\int_1^{\mathrm{R_{ch}}}(x-1)^4(x^2+4x+5)\frac{g_{11}(yR)^{1/2}y^2dy}{g_{00}(yR)^{9/2}}\,\ ,
\end{align}
where $g_{00}(r)$ and $g_{11}(r)$ are defined in (\ref{metricKerr}).
The function $\mathcal{F}_J\left(\frac{M}{R}\right)$ reduces to  $\mathcal{F}\left(\frac{M}{R}\right)$ for $J=0$ (Eq. (\ref{fMRSchw})).

\begin{figure}
    \centering
    \includegraphics[width=0.9\textwidth]{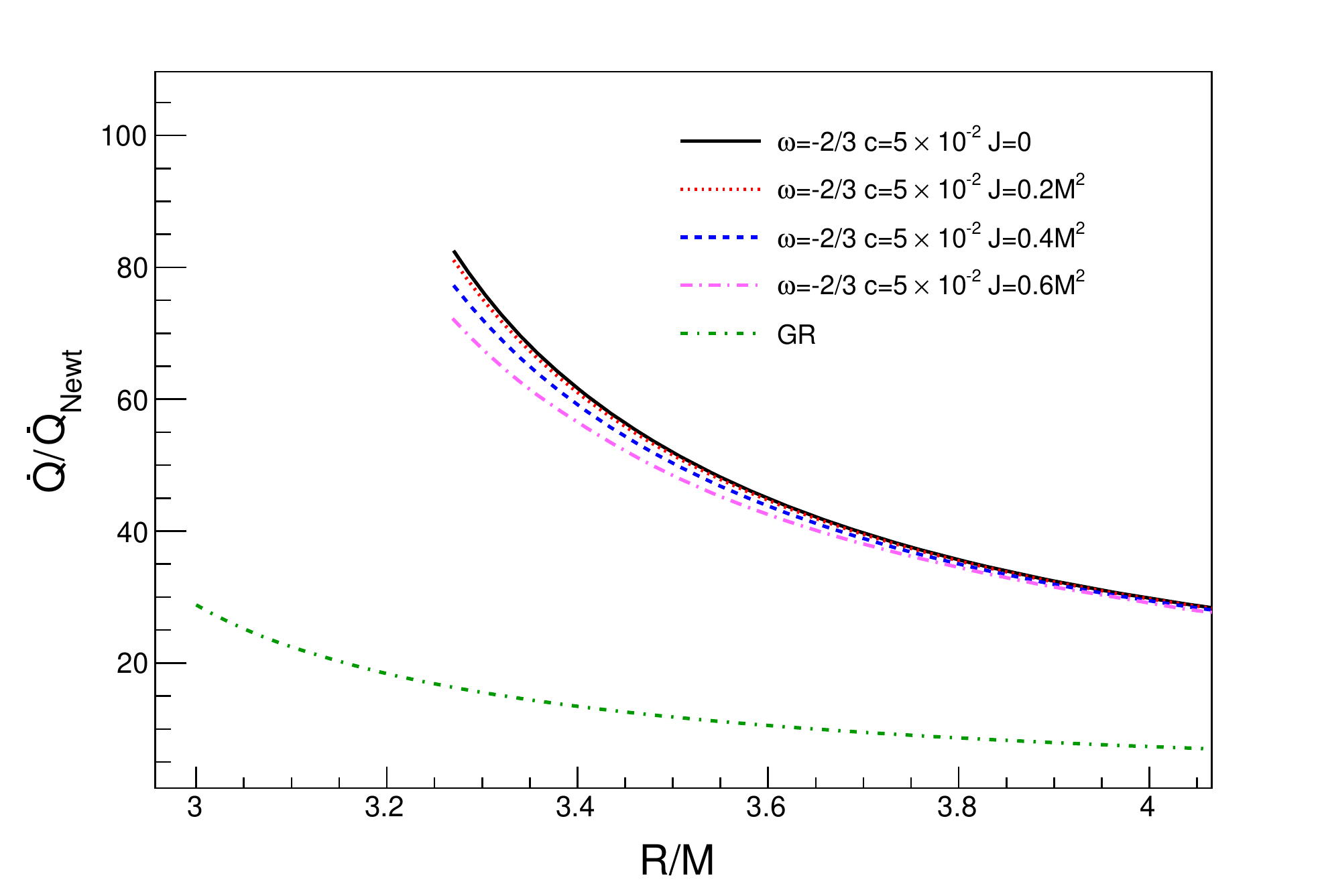}
    \caption{Ratio of total energy deposition $\dot{Q}$ for $\omega=-2/3$ and $c=0.05$ to total Newtonian energy deposition $\dot{Q}_{\mathrm{Newt}}$ for four values of the parameter $J$. The green curve shows the GR energy deposition for comparison.}
    \label{rot}
\end{figure}

As in the previous Section, we write the energy deposition due to the neutrino annihilation process, Eq. (\ref{contoQ}), in terms of the the ratio $\dot{Q}/\dot{Q}_{\mathrm{Newt}}$, where now for rotating source reads
 \begin{equation}\label{QratioJ}
\frac{\dot{Q}}{\dot{Q}_{\mathrm{Newt}}}=
\mathcal{F}_J\left(\frac{M}{R}\right)\,.
\end{equation}
In Figure ~\ref{rot} we report the ratio $\dot{Q}/\dot{Q}_{\mathrm{Newt}}$ vs $R/M$ in the case of a slowly rotating gravitational source. The plot shows that the angular momentum $J$ induces only slight modifications with respect to the non-rotating case, Moreover, we note that higher values of the angular momentum $J$ reduces the ratio $\dot{Q}/\dot{Q}_{\mathrm{Newt}}$, an effect due to the fact that the $J$-term change $\cos\theta_r$ as in Eq.~(\ref{cosrJ}) (we have checked that this is true for both positive and negative angular momentum of the source). 
According to these considerations, for a slowly rotating gravitational source, we get almost the same results of the non-rotational case.
The extension to the case of high values of the angular momentum $J$ is of interest, although the solutions of the Einstein field equation turned to be extremely involved.

\section{Conclusion}
In this paper, we have analyzed the neutrino pair annihilation $\nu\bar{\nu}\rightarrow e^+ e^-$
in the framework of BH surrounding by a quintessence field. We have shown that in the allowed range of parameters of the quintessence model a shift of the photosphere radius occurs, as compared to the one computed in GR, and a consequent enhancement of the emitted energy rate ratio  $\dot{Q}/\dot{Q}_{\mathrm{Newt}}$. Such an enhancement could be relevant for the generation of GRBs in close neutron star binary merging, for which neutrino pairs annihilation has been proposed as a possible source, and can be used to find constraints on the parameter $c$ of the model (see Figs.~\ref{cont1} and \ref{cont2}).

The main purpose of the paper is hence to  analyzed the case in which the space-time around a gravitational source is described by a Schwarzschild-like geometry modified by the presence of a quintessence field, and to use GRB to establish constraints on the model's parameters. As stressed in \cite{Salmonson:1999es}, the physics near the surface of a hot ($T \sim$ MeV) neutron star is not trivial, since interactions among particles (baryons, leptons, photons and neutrinos) are involved and play an important role in energy conversion and transport. Here we focused on the process $\nu  \overline{\nu} \rightarrow e^+  e^-$, that represents only a small component with respect to all mentioned processes. Moreover, the possible interaction of particles with the quintessence field has been neglected.
According to \cite{Salmonson:1999es}, results here derived could be non-trivial near the surface of a hot neutron star, such as Supernovae
and merged neutron star binaries. Moreover, the mechanism for the generation of GRBs from the merging of a neutron stars binary system sourced by the annihilation of neutrino-antineutrino pairs, owing to the deformation of the space-time geometry, may provide the needed energy close to the energy liberated during GRBs. 
The analysis performed in this paper refers to a static spherically symmetric quintessence surrounding a gravitational source. We have also investigated the case corresponding to a slowly rotating gravitational source. The model can be generalized to include not only rotational effects in strong gravitational regimes  \cite{Toshmatov:2015npp,Ghosh:2015ovj}, but also the effects of (nonlinear) magnetic fields (see for example \cite{MosqueraCuesta:2017iln,Jamil:2014rsa}). In these scenarios, the presence of the quintessence field in the
surrounding of a Black Holes could significantly affect the gravitational influence on the neutrino pair annihilation and increases the neutrino luminosity factor, needed to efficiently generate the GRBs. These studies will be faced elsewhere.  

A final comment is in order. Typically, it is expected that there is a connection between (short) GRBs and Gravitational Waves (GWs). Their correlation is important because would not only provide new astrophysical insight about the transient phenomena but also it would confirm, in the case of a binary neutron star merger, where this event occurred (being at the same time an indirect proof of that it actually happened). However, there are recent results of the searching for GWs associated with GRBs detected during the first observing run of the Advanced Laser Interferometer Gravitational-Wave Observatory (LIGO), that indicate that there is no evidence of a GW signal for the 41 GRBs analyzed \cite{Abbott:2016cjt}. According to these results (i.e. there could not be a strict correlation among GWs, merging of neutron stars and GRBs), alternative mechanisms for the generation of GRBs could provide  in future new scenarios for understanding the GRBs physics.


\acknowledgments
We thank the referee for comments that improved the paper. The work of G.L. and L.M. is supported by the Italian Istituto Nazionale di Fisica Nucleare (INFN) through the ``QGSKY'' project and by Ministero dell'Istruzione, Universit\`a e Ricerca (MIUR).
Part of these activities (G.L.) was performed within the scientific working groups of the THESEUS Consortium (http://www.isdc.unige.ch/theseus/ -  THESEUS is a space astrophysics mission concept under Phase A assessment study by ESA as candidate M5 mission within the Cosmic VIsion programme).
The computational work has been executed on the IT resources of the ReCaS-Bari data center, which have been made available by two projects financed by the MIUR (Italian Ministry for Education, University and Research) in the "PON Ricerca e Competitività 2007-2013" Program: ReCaS (Azione I - Interventi di rafforzamento strutturale, PONa3\_00052, Avviso 254/Ric) and PRISMA (Asse II - Sostegno all'innovazione, PON04a2A).

\appendix

\section{Escape probability for deposited energy}
\label{appendice}
To determine the escape probability for the deposited energy, we have to use the photon equation of motion (we follow \cite{Shapiro:1983du})
\begin{equation}
    \left(\frac{dr}{d\tau}\right)^2=\frac{1}{b^2}-V_{\mathrm{eff}}(r)
\end{equation}
where $V_{\mathrm{eff}}$ is the effective potential for a photon. In GR it has the form:
\begin{equation}
    V_{\mathrm{eff}}=\frac{1}{r^2}\left(1-\frac{2M}{r}\right) \,\ .
\end{equation}
This quantity has a maximum at $r=3M$ which define the critical impact parameter:
\begin{equation}
    b_c=3\sqrt{3}M \,\ ,
\end{equation}
which is a distinguish for incoming particle between absorption and scattering. Therefore, for $r\geq 3M$, a particle is emitted if $v^r>0$ or $b>b_c$, with $v^r$ the radial velocity measured by a static observer.\\
It is possible to define $v^r=\cos\phi$ and $v^{\phi}=\sin\phi$ and one can write that:
\begin{equation}
    v^{\phi}=\frac{b}{r}\sqrt{\left(1-\frac{2M}{r}\right)} \,\ .
\end{equation}
Therefore, finally, one can write that for $r>3M$ a particle escape if:
\begin{equation}
    0<\phi<\frac{\pi}{2}+\arcsin{\frac{b_c}{r}\sqrt{\left(1-\frac{2M}{r}\right)}}=\frac{\pi}{2}+\arcsin{\frac{3\sqrt{3}M}{r}\sqrt{\left(1-\frac{2M}{r}\right)}} \,\ .
    \label{angolo}
\end{equation}
This are the bounds we have to take in consideration while doing angular integration, that lead to the probability: 
\begin{equation}
    C(r)=\frac{1}{2}\left[1+\sqrt{1-\left(\frac{3\sqrt{3}M}{r}\left(1-\frac{2M}{r}\right)\right)^2}\right] \,\ .
\end{equation}
As it can be seen, the probability is always larger than $1/2$. in the case of modified theories of gravity, we will have a change in $V_{\mathrm{eff}}$, but due to Eq.~(\ref{angolo}), we will always have that $C(r)\geq 1/2$.

\section{Static spherically symmetric black holes surrounded by quintessence}

In this Appendix we shortly review the derivation of Eq. (\ref{metric}) corresponding to a static spherically-symmetric black hole surrounded by quintessence in $d$-dimensions. We strictly follow the Chen-Wang-Su paper \cite{Chen:2008ra}. The geometry of a static black hole is described by
\begin{eqnarray}\label{AGae1}
ds^2=e^{\nu(r)}dt^2-e^{\lambda(r)}dr^2-r^2d\theta^2_1-r^2\sin^2{\theta_1}d\theta^2_2-\cdots-r^2\sin^2{\theta_1}
\cdots\sin^2{\theta_{d-3}}d\theta^2_{d-2}\,.
\end{eqnarray}
For a static spherically symmetric configuration, the quintessence energy density and pressure can be written in the form 
\cite{Kiselev:2002dx}
\begin{equation}\label{Tcomps}
T^{\;0}_0 = A(r), \quad T^{\;j}_0=0,\quad T^{\;j}_i=C(r)r_i r^j+B(r)\delta^{\;j}_i,
\end{equation}
from which, averaging over the angles of isotropic state $\langle r_ir^j\rangle =\frac{1}{d-1}r_k r^k \delta^{\;j}_i$, one gets
\begin{equation}
\langle T^{\;j}_i\rangle = D(r)\delta^{\;j}_i,\quad D(r)=-\frac{1}{d-1}C(r)r^2+B(r).
\end{equation}
The quintessence is characterized by the condition $p =-\omega_q \rho$, that is $D(r)=-\omega_q A(r)$.  
As in \cite{Kiselev:2002dx,Chen:2008ra}  one considers $C(r) \propto B(r)$, so that the (exact) solutions with charged or not charged black holes are possible, as well as the generalization to asymptotically flat or de Sitter spacetime.  The constant coefficient between $C(r)$  and $B(r)$  is defined by the additivity and linearity condition\footnote{Such a condition allows to obtain the correct limits for the well-known cases of charged black holes ($w_q = 1/3$), dust matter ($w_q=0$), and quintessence ($w_q = -1$)  \cite{Kiselev:2002dx}.}.

The metric (\ref{AGae1}) allows to write down the Einstein field equations
\begin{eqnarray}
T^{\;0}_0 &=& \frac{d-2}{4}\left[-e^{-\lambda}\bigg(\frac{d-3}{r^2}-\frac{\lambda'}{r}\bigg)+\frac{d-3}{r^2}\right]\,,\nonumber \\
T^{\;r}_r&=&\frac{d-2}{4}\left[-e^{-\lambda}\bigg(\frac{d-3}{r^2}+\frac{\nu'}{r}\bigg)+\frac{d-3}{r^2}\right]\,, \nonumber
\\
T^{\;\theta_1}_{\theta_1} &=& T^{\;\theta_2}_{\theta_2}=\cdots T^{\;\theta_{d-2}}_{\theta_{d-2}}\nonumber\\
&=&-\frac{e^{-\lambda}}{4}\bigg[\nu''+\frac{\nu'^2}{2}-\frac{\lambda'\nu'}{2}
+\frac{(d-3)(\nu'-\lambda')}{r}+\frac{(d-3)(d-4)}{r^2}\bigg]+\frac{(d-3)(d-4)}{2r^2}\,, \nonumber
\end{eqnarray}
where ${}^\prime\equiv \partial/\partial r$. 
Using the appropriate general expression of the energy-momentum tensor of quintessence in the $d-$dimensional spherically-symmetric spacetime\footnote{The spatial part of the energy-momentum tensor is (proportionally) related to the time component with the arbitrary parameter $B$ which does depend on the internal structure of quintessence} 
\begin{equation}\label{A2} 
T^{\;0}_0=\rho_q(r),\quad T^{\;j}_i=\rho_q(r)\alpha
\left\{-[1+(d-1)B]\frac{r_ir^j}{r_k r^k}+B\delta^{\;j}_i\right\},
\end{equation}
the equation of state $p_q=\omega_q\rho_q$, and taking the isotropic average over the angles for which 
$\langle T^{\;j}_i\rangle =-\rho_q(r)\frac{\alpha}{d-1}\delta^{\;j}_i\equiv -p_q\delta^{\;j}_i$,
one gets $\omega_q=\frac{\alpha}{d-1}$.
For quintessence one has $-1<\omega_q<0$ and, consequently, $-(d-1)<\alpha<0$. Moreover, requiring
$T^{\;0}_0=T^{\;r}_r$ (this defines the principle of additivity and
linearity \cite{Kiselev:2002dx}), which implies $\lambda+\nu=0$
and substituting $\lambda=-\ln{f}$,
one obtains 
\begin{eqnarray}
T^{\;0}_0&=&T^{\;r}_r=-\frac{d-2}{4r^2}[rf'+(d-3)(f-1)],\label{Appt1}\\
T^{\;\theta_1}_{\theta_1}&=&T^{\;\theta_2}_{\theta_2}=\cdots
T^{\;\theta_{d-2}}_{\theta_{d-2}}=
-\frac{1}{4r^2}[r^2f''+2(d-3)rf'+(d-4)(d-3)(f-1)]\label{Appt2}.
\end{eqnarray}
From equations (\ref{A2}) and (\ref{Appt1}) it immediately follows $B=-\frac{\alpha+1}{\alpha (d-2)}=-\frac{(d-1)\omega_q+1}{(d-1)(d-2)\omega_q}$.
The components of the energy-momentum tensor (\ref{A2}) read
\begin{eqnarray}
T^{\;0}_0&=&T^{\;r}_r=\rho_q,\label{t3} \\
T^{\;\theta_1}_{\theta_1}&=&T^{\;\theta_2}_{\theta_2}=\cdots =
T^{\;\theta_{d-2}}_{\theta_{d-2}}=-\frac{1}{d-2}\rho_q[(d-1)\omega_q+1]\label{t4}\,,
\end{eqnarray}
while the combination of Eqs. (\ref{Appt1})-(\ref{t4}) leads to the differential equation for $f$
\begin{eqnarray}\label{de1}
r^2f''+[(d-1)\omega_q+2d-5]rf'+(d-3)[(d-1)\omega_q+d-3](f-1)=0\,,
\end{eqnarray}
whose general solution is of the form
\begin{eqnarray}
f=1-\frac{r_g}{r^{d-3}}+\frac{c_1}{r^{(d-1)\omega_q+d-3}},
\end{eqnarray}
where $r_g$ and $c_1$ are normalization factors. For $c_1=0$ one recovers
the usual $d$-dimensional Schwarzschild
solution.  The energy density $\rho_q$ for quintessence is given by
$\rho_q=\frac{c_1\omega_q(d-1)(d-2)}{4r^{(d-1)(\omega_q+1)}}$.
Since $\rho_q>0$ and $\omega_q\leq 0$, it follows  
$c_1<0$. Taking $r_g=2M$ and $c_1=-c$, the metric of the $d$-dimensional
spherically symmetric black hole surrounded by quintessence reads
\begin{eqnarray}\label{A3}
ds^2=\bigg[1-\frac{2M}{r^{d-3}}-\frac{c}{r^{(d-1)\omega_q+d-3}}\bigg]dt^2
-\frac{dr^2}{\displaystyle{1-\frac{2M}{r^{d-3}}-\frac{c}{r^{(d-1)\omega_q+d-3}}}}
-r^2d\Omega_{d-2}.
\end{eqnarray}
The metric (\ref{A3}) does depend on the quintessence state parameter $\omega_q$. 
For $d=4$, Eq. (\ref{A3}) reduces to (\ref{metric}) and (\ref{frmetric}).

\bibliographystyle{JHEP.bst}  
\bibliography{main.bib}    

\end{document}